\documentclass[prd, twocolumn,reprint]{revtex4-1}
\usepackage{overpic}
\usepackage{graphicx}
\usepackage{epsfig}
\usepackage{dcolumn}
\usepackage{bm}
\usepackage{rotating}
\usepackage{multirow}
\usepackage{booktabs}
\usepackage{appendix}
\usepackage{subfigure}
\usepackage{hhline}
\usepackage{amssymb}
\usepackage{lineno}
\usepackage{amsmath}
\usepackage{hyperref}
\usepackage{cleveref}
\usepackage{mathrsfs}
\usepackage{verbatim}
\usepackage{amssymb}
\usepackage{amsfonts}
\usepackage{amsbsy}
\usepackage{float}
\usepackage{lipsum}
\usepackage{float}
\hypersetup{colorlinks,linkcolor=blue,anchorcolor=blue,citecolor=blue}

\let\oldequation\equation
\let\oldendequation\endequation

\renewenvironment{equation}
  {\linenomathNonumbers\oldequation}
  {\oldendequation\endlinenomath}

\def \miss2{M_{\rm miss}^{2}}

\def \romanOne   {\uppercase\expandafter{\romannumeral1}}
\def \romanTwo   {\uppercase\expandafter{\romannumeral2}}
\def \romanThree {\uppercase\expandafter{\romannumeral3}}
\def \romanFour  {\uppercase\expandafter{\romannumeral4}}
\def \romanFive  {\uppercase\expandafter{\romannumeral5}}
\def \romanSix   {\uppercase\expandafter{\romannumeral6}}
\def \romanSeven {\uppercase\expandafter{\romannumeral7}}
\def \romanEight {\uppercase\expandafter{\romannumeral8}}
\def \romanNine {\uppercase\expandafter{\romannumeral9}}

\newcommand{\lambdacp}{\Lambda_{c}^{+}}

\newcommand{\sigmode}[1]{
	\ifnum#1=1
	\lambdacp \rightarrow n K_{S}^{0} \pi^{+}
	\else
	\ifnum#1=2
	\lambdacp \rightarrow n K_{S}^{0} K^{+}
	\fi
	\fi
}

\begin{document}

\graphicspath{{figure/}}
\DeclareGraphicsExtensions{.eps,.png,.ps}

 \newcommand{\BESIIIorcid}[1]{\href{https://orcid.org/#1}{\hspace*{0.1em}\raisebox{-0.45ex}{\includegraphics[width=1em]{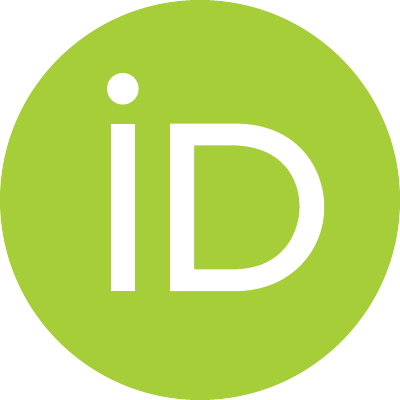}}}}

\title{\boldmath First observation of the $\eta_{c}\to\Xi^{0} \bar{\Xi}^{0}$ decay}

\author{
\begin{small}
\begin{center}
M.~Ablikim$^{1}$\BESIIIorcid{0000-0002-3935-619X},
M.~N.~Achasov$^{4,d}$\BESIIIorcid{0000-0002-9400-8622},
P.~Adlarson$^{81}$\BESIIIorcid{0000-0001-6280-3851},
X.~C.~Ai$^{87}$\BESIIIorcid{0000-0003-3856-2415},
C.~S.~Akondi$^{31A,31B}$\BESIIIorcid{0000-0001-6303-5217},
R.~Aliberti$^{39}$\BESIIIorcid{0000-0003-3500-4012},
A.~Amoroso$^{80A,80C}$\BESIIIorcid{0000-0002-3095-8610},
Q.~An$^{77,64,\dagger}$,
Y.~H.~An$^{87}$\BESIIIorcid{0009-0008-3419-0849},
Y.~Bai$^{62}$\BESIIIorcid{0000-0001-6593-5665}
% ...
\\
\vspace{0.2cm}
(BESIII Collaboration)\\
\vspace{0.2cm} {\it
$^{1}$ Institute of High Energy Physics, Beijing 100049, People's Republic of China\\
% ...
$^{4}$ Budker Institute of Nuclear Physics SB RAS (BINP), Novosibirsk 630090, Russia\\
% ...
$^{31}$ INFN Sezione di Ferrara, (A)INFN Sezione di Ferrara, I-44122, Ferrara, Italy; (B)University of Ferrara, I-44122, Ferrara, Italy\\
% ...
$^{39}$ Johannes Gutenberg University of Mainz, Johann-Joachim-Becher-Weg 45, D-55099 Mainz, Germany\\
% ...
$^{62}$ Southeast University, Nanjing 211100, People's Republic of China\\
% ...
$^{64}$ State Key Laboratory of Particle Detection and Electronics, Beijing 100049, Hefei 230026, People's Republic of China\\
% ...
$^{77}$ University of Science and Technology of China, Hefei 230026, People's Republic of China\\
% ...
$^{80}$ University of Turin and INFN, (A)University of Turin, I-10125, Turin, Italy; (B)University of Eastern Piedmont, I-15121, Alessandria, Italy; (C)INFN, I-10125, Turin, Italy\\
$^{81}$ Uppsala University, Box 516, SE-75120 Uppsala, Sweden\\
% ...
$^{87}$ Zhengzhou University, Zhengzhou 450001, People's Republic of China\\
\vspace{0.2cm}
$^{\dagger}$ Deceased\\
% ...
$^{d}$ Also at the Novosibirsk State University, Novosibirsk, 630090, Russia\\
% ...
}
\end{center}
\vspace{0.4cm}    
\end{small}
} 

\begin{abstract}
Using $(10087\pm 44)\times 10^6$ $J/\psi$  events collected with the BESIII detector at the BEPCII collider, we report the first observation of the decay $\eta_{c} \to \Xi^{0} \bar{\Xi}^{0}$. The interference between $J/\psi\to\gamma\eta_c\to\gamma\Xi^0\bar{\Xi}^0$ and $J/\psi\to\gamma\Xi^0\bar{\Xi}^0|_{\rm non-resonance}$  is considered in the $\Xi^0 \bar{\Xi}^0$ mass spectrum fits. The branching fractions are measured to be $\mathcal{B}(\eta_c \to \Xi^0 \bar{\Xi}^0)=(1.33 \pm 0.03 \pm 0.18) \times 10^{-3}$ for the constructive interference and $\mathcal{B}(\eta_c \to \Xi^0 \bar{\Xi}^0)=(1.63 \pm 0.04 \pm 0.21) \times 10^{-3}$ for the destructive interference, where the first uncertainties are statistical and the second are systematic.

\end{abstract}

\maketitle
\oddsidemargin -0.2cm
\evensidemargin -0.2cm

\section{\boldmath Introduction}

The hadronic decays of charmonium provide an ideal platform for testing the interplay between perturbative and non-perturbative Quantum Chromodynamics (QCD). The charmonium mass scale lies in a transition region where QCD dynamics evolve from the perturbative regime at high energies to the non-perturbative regime at low energies.

The complexity of QCD in the charmonium-mass region is highlighted by significant discrepancies between perturbative Quantum Chromodynamics (pQCD) calculations and experimental measurements. A prime example is the long-standing puzzle associated with the pQCD helicity selection rule (HSR)~\cite{pQCD_1,pQCD_2,pQCD_3}. Derived from the quark helicity conservation principle in point-like interactions, the HSR predicts that helicity-violating decays of the $0^{-+}$ pseudoscalar charmonium state $\eta_c$ into baryon--antibaryon ($B\bar{B}$) final states should be forbidden~\cite{HSR}. However, experimental measurements consistently contradict this prediction, with measured branching fractions that are comparable to those of $J/\psi$ decays~\cite{BESIII:2019eyx,BESIII:2012xdg,BESIII:2012gcb,PDG}. These discrepancies suggest the presence of previously unaccounted-for mechanisms in pQCD, making HSR violation a critical issue to be addressed.

Several theoretical frameworks have been proposed to account for this phenomenon. Models such as the quark--diquark model~\cite{Anselmino:1987du,Anselmino:1991mc}, charmonium--glueball mixing~\cite{Anselmino:1993yg}, the quark-pair creation model~\cite{Ping:2004sh}, and the color-octet mechanism~\cite{Wong:1999hc} attribute HSR violation to the internal structure of charmonium. These approaches allow otherwise forbidden decays to occur, yielding nonzero branching fractions that can be comparable to those of HSR-allowed decays. Other theoretical studies suggest that such transitions may arise from non-perturbative corrections~\cite{Anselmino:1992jd,Murgia:1996bh}. Meanwhile, the Intermediate Meson Loop (IML) model~\cite{Zhang:2009kr,Liu:2009vv} provides a dynamical mechanism for evading the HSR by introducing long-range effects through virtual charmed-meson loops. In this model, the decay proceeds not as a point-like interaction but via a two-step process: the $\eta_c$ first decays into a virtual charmed-meson pair~\cite{Liu:2010um} (e.g.\ $D\bar{D}$), which subsequently forms the final-state baryon--antibaryon pair through the exchange of light mesons. This long-range mechanism effectively circumvents the assumptions underlying the pQCD HSR and can result in sizeable branching fractions for otherwise forbidden decays.

Despite their qualitative success in explaining HSR violation, these theoretical frameworks generally suffer from large quantitative uncertainties in their predictions of branching fractions. Consequently, experimental measurements of additional helicity-forbidden decay channels, particularly those not yet observed, are essential for testing the validity and limitations of existing models.

To further probe these unresolved issues, a measurement of the previously unobserved decay $\eta_c \to \Xi^{0}\bar{\Xi}^{0}$ is of particular importance, as it is the isospin partner of $\eta_c \to \Xi^{-}\bar{\Xi}^{+}$. Although the $\eta_c$ cannot be directly produced in $e^{+}e^{-}$ annihilation due to quantum number conservation, it is copiously produced via radiative decays of the $J/\psi$ resonance. Therefore, the large data sample of $(10087 \pm 44)\times10^{6}$ $J/\psi$ events collected by the BESIII experiment~\cite{BESIII:2021cxx} enables such a measurement for the first time. In this paper, we report the first observation of this decay and a measurement of its branching fraction, using $\eta_c$ mesons produced through the radiative process $J/\psi \to \gamma\eta_c$.

\section{\boldmath BESIII Experiment and Monte Carlo Simulation}

The BESIII detector~\cite{Ablikim:2009aa} records symmetric $e^+e^-$ collisions provided by the BEPCII storage ring~\cite{Yu:2016cof}, which operates with a peak luminosity of $1.1 \times 10^{33}\;\text{cm}^{-2}\text{s}^{-1}$ in the center of mass energy range from 1.84 to 4.95 GeV. BESIII has collected large data samples in this energy region~\cite{Ablikim:2019hff,EcmsMea,EventFilter}. The cylindrical core of the BESIII detector covers 93\% of the full solid angle and consists of a helium-based multilayer drift chamber (MDC), a plastic scintillator time-of-flight system (TOF), and a CsI(Tl) electromagnetic calorimeter (EMC), which are all enclosed in a superconducting solenoidal magnet providing a 1.0~T magnetic field. The magnetic field was 0.9~T in 2012, which affects 13\% of the total $J/\psi$ data. The solenoid is supported by an
octagonal flux-return yoke with resistive plate counter muon
identification modules interleaved with steel. The charged-particle momentum resolution at $1~{\rm GeV}/c$ is $0.5\%$, and the 
${\rm d}E/{\rm d}x$ resolution is $6\%$ for electrons from Bhabha scattering. The EMC measures photon energies with a resolution of $2.5\%$ ($5\%$) at $1$~GeV in the barrel (end cap) region. The time resolution in the plastic scintillator TOF barrel region is 68~ps, while that in the end cap region was 110~ps. The end cap TOF system was upgraded in 2015 using multigap resistive plate chamber technology, providing a time resolution of 60~ps, which benefits ~87\% of the data used in this analysis~\cite{etof}.

Monte Carlo (MC) simulated data samples produced with a {\sc
geant4}-based~\cite{geant4} software package, which
includes the geometric description of the BESIII detector and the
detector response, are used to determine detection efficiencies
and to estimate backgrounds. The simulation models the beam
energy spread and initial state radiation in the $e^+e^-$
annihilations with the generator {\sc
kkmc}~\cite{ref:kkmc}. The inclusive MC sample includes both the production of the $J/\psi$
resonance and the continuum processes incorporated in {\sc
kkmc}.
All particle decays are modelled with {\sc
evtgen}~\cite{ref:evtgen} using branching fractions 
either taken from the
Particle Data Group (PDG)~\cite{PDG}, when available,
or otherwise estimated with {\sc lundcharm}~\cite{ref:lundcharm}.
Final state radiation from charged final state particles is incorporated using the {\sc photos} package~\cite{photos2}. The potential background events in the inclusive MC are examined using the event-type analysis tool TopoAna~\cite{Zhou:2020ksj}. To assess the detection efficiency, signal MC events are generated following the decay chain $J/\psi\to\gamma\eta_{c},\eta_{c} \to \Xi^{0} \bar{\Xi}^{0}, \Xi^{0} \to \Lambda \pi^0, \bar{\Xi}^{0} \to \bar{\Lambda} \pi^0, \Lambda \to p \pi^{-}, \bar{\Lambda} \to \bar{p} \pi^{+}$, and $\pi^0 \to \gamma \gamma$. The decay $J/\psi\to\gamma\eta_{c}$ is generated using JPE~\cite{ref:lundcharm}, where the JPE is a Generator model constructed for a vector particle decaying into a photon plus a pseudoscalar particle, and the subsequent decays are simulated based on the phase space (PHSP) model.

\section{\boldmath Event Selection}
\label{sec:selection}

We reconstruct the signal candidates through the decay chain described in the previous section, resulting in the final state including $\gamma\gamma\gamma\gamma\gamma p \bar{p} \pi^+ \pi^- $.

Events with at least four charged tracks with total charge zero are selected.
Charged tracks detected in the MDC are required to satisfy the polar angle ($\theta$) condition $|\cos\theta| < 0.93$, where $\theta$ is defined with respect to the $z$-axis, the symmetry axis of the MDC. 
Given the relatively long lifetimes of $\Lambda$  and $\Xi^0$, no specific vertex requirement is imposed on the charged tracks decaying from them.  
MC studies indicate that the proton and pion from $\Lambda$ decay are well separated kinematically since the proton carries most of the energy. The charged tracks with momentum above 0.247 GeV$/c$ are supposed to be protons, while those below 0.247 GeV$/c$ are pions.
The $\Lambda(\bar{\Lambda})$ candidates are reconstructed from 
$p\pi^{-}(\bar{p}\pi^{+})$ combinations. They are constrained to originate from a common vertex, with those having 
$\chi^{2}<200$ kept for further analysis. Furthermore, the mass window of $p\pi^{-}(\bar{p}\pi^{+})$ is set to be $(1.111,1.120)~\text{GeV}/c^{2}$.

Photon candidates are identified using isolated showers in the EMC and required to contain at least five photons per event.  The deposited energy of each shower must be more than 25~MeV in the barrel region ($|\cos \theta|< 0.80$) and more than 50~MeV in the end cap region ($0.86 <|\cos \theta|< 0.92$).
To exclude showers that originate from
charged tracks,
the angles subtended by the EMC shower and the position of the closest charged track at the EMC
must be greater than 10 degrees as measured from the interaction point. A more stringent requirement of 20 degrees between the photon candidate and anti-proton is applied to exclude the photons from anti-proton annihilation.
To suppress electronic noise and showers unrelated to the event, the difference between the EMC time and the event start time is required to be within $[0, 700]$ ns. 
The $\pi^0$ candidate is reconstructed from a photon pair with $M_{\gamma\gamma}\in (0.098,0.165)$ GeV/$c^2$ following the published study~\cite{BESIII:2023drj}. A one-constraint kinematic fit is performed on the photon pairs by constraining their invariant mass to the nominal $\pi^0$ mass~\cite{PDG}.

A four-constraint (4C) kinematic fit is performed on the $\gamma\Lambda\bar{\Lambda}\pi^0\pi^0$ hypothesis, constraining the total reconstructed four-momentum to that of the initial $e^+e^-$. When the number of photons exceeds five, all possible combinations are tested, and the combination yielding the minimum $\chi^2$ of 4C kinematic fit ($\chi^2_{\rm{4C}}$) is retained for further study. To enhance background suppression while preserving more signal events, a Figure of Merit (FOM) method~\cite{PDG} is employed to optimize the selection criterion of $\chi^2_{\rm{4C}}$ by maximizing the $S/\sqrt{S+B}$ value, where $S$ and $B$ represent the numbers of surviving signal and background events in the inclusive MC sample, respectively. Following the optimization result, candidate events are required to meet the condition $\chi^2_{\rm{4C}}<30$ and the background events violating four-momentum conservation are removed. To further suppress backgrounds with photon counts deviating from five in the final states, we require that the $\chi^2_{\rm{4C}}$ with the   $\Lambda\bar{\Lambda}\pi^0\pi^0$ hypothesis should be more than 200. This requirement specifically suppresses the $J/\psi \to \Xi^{0} \bar{\Xi}^{0}$ background without substantial efficiency loss.

\begin{figure}[hpbt]
    \centering
      \includegraphics[width=0.75\linewidth]{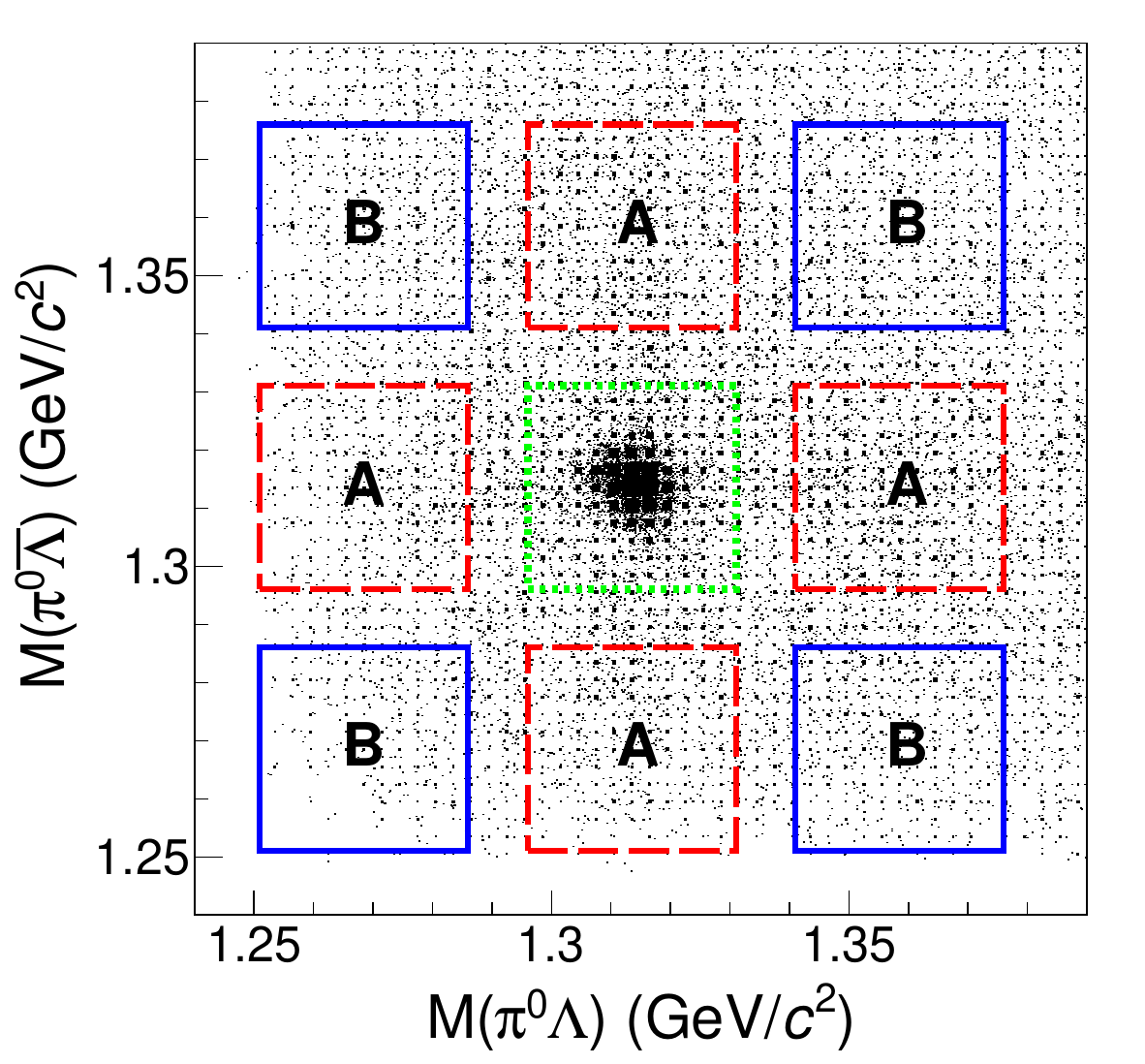}
      \caption{ Distribution of the  invariant mass of $\pi^0\Lambda$ versus $\pi^0\bar\Lambda$ for data, showing the signal (green box) and sideband regions A and B (red and blue boxes, respectively).}
    \label{fig:sideband_B_C}
\end{figure}

In the study of $\eta_{c} \to \Xi^{0} \bar{\Xi}^{0}$ decay, since there are two $\pi^0$ candidates (denote as $\pi^0_1, \pi^0_2)$ per event, there will be two possible combinations of $\Lambda(\bar{\Lambda})$ and $\pi^0$. The combination which minimizes the quantity $(m_{\pi^0_1\Lambda}-M_{\Xi}^0)^2+(m_{\pi^0_2\bar{\Lambda}}-M_{\bar{\Xi}^0})^2$ is kept, where $m_{\pi^0_1\Lambda}$ and $m_{\pi^0_2\bar{\Lambda}}$ are the invariant masses of $\pi^0_1\Lambda$ and $\pi^0_2\bar{\Lambda}$, respectively, and $M_{\Xi^0}$($M_{\bar{\Xi}^0}$) is the known mass of $\Xi^0$($\bar{\Xi}^0$)\cite{PDG}. Additionally,  the invariant masses of $\pi^{0}\Lambda$ and $\pi^{0}\bar{\Lambda}$ pairs are required to be within $(1.296, 1.331)$ GeV/$c^2$. The mass window is determined to be ($\mu\pm 3\sigma$) through MC study, where $\mu$ represents the mean value obtained from the Gaussian fit and $\sigma$ represents the mass resolution. The distribution of $M(\pi^0\Lambda)$ versus $M(\pi^0\bar\Lambda)$ from data is shown in Fig.~\ref{fig:sideband_B_C}, where a clear $\Xi^{0} \bar{\Xi}^{0}$ signal can be seen as indicated by the central green box.

\begin{figure*}[btp]
  \centering
    \includegraphics[width=0.45\linewidth]{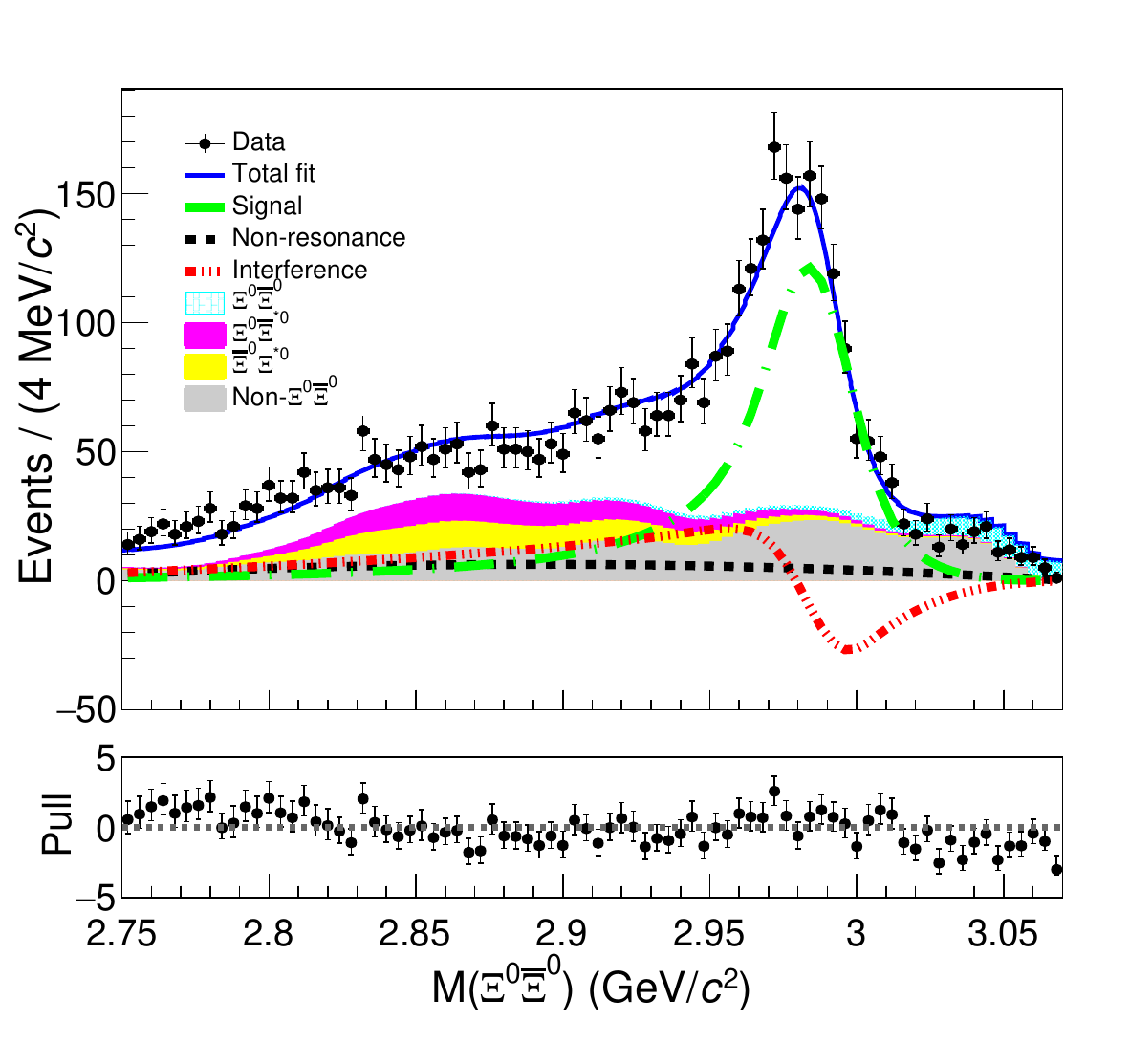}
    \includegraphics[width=0.45\linewidth]{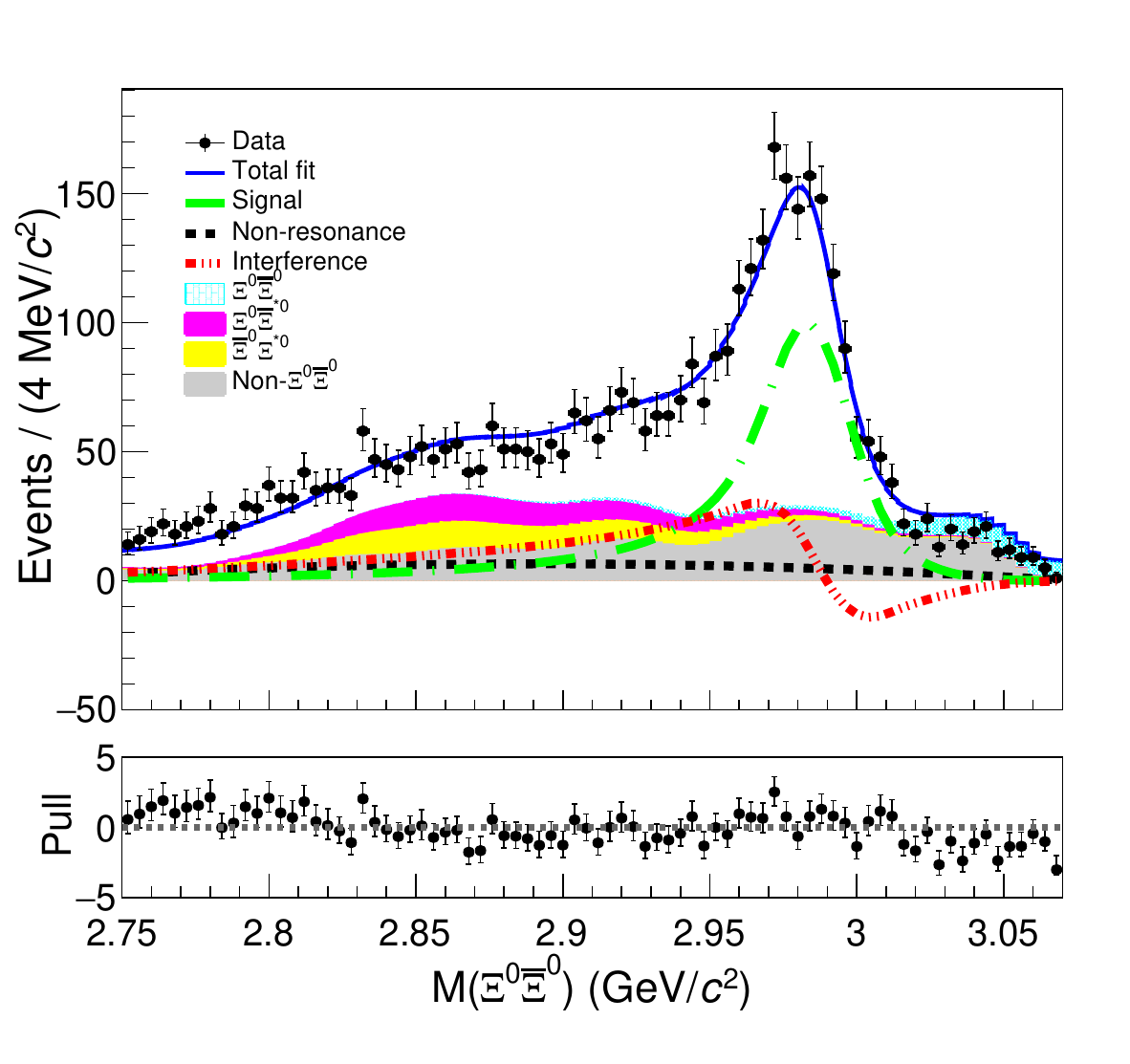}
    \caption{Fit to the $M(\Xi^0\bar{\Xi}^0)$ distribution for the (left) destructive and (right) constructive interference solutions. Data (black points) are overlaid with the total fit (blue line). Key components: signal (green dashed, $J/\psi \to \gamma\eta_c \to \gamma\Xi^0\bar{\Xi}^0$), non-resonance part (black dashed, $J/\psi \to \gamma\Xi^0\bar{\Xi}^0$), interference (red dotted), $J/\psi \to \Xi^{0} \bar{\Xi}^{0}$ (azure fill), $J/\psi \to \Xi^{*0} \bar{\Xi}^{0}$/$J/\psi \to \Xi^{0} \bar{\Xi}^{*0}$ (yellow/violet fill), and non-$\Xi^{0} \bar{\Xi}^{0}$ background (gray fill). The pull distributions are shown in the lower panel.}
  \label{fig:conclusion_result}
\end{figure*}

\section{\boldmath Background Studies}\label{sec:background_study}
 
The background components are studied with the $J/\psi$ inclusive MC sample using TopoAna. After applying the same selection criteria as for data, the background can be classified into two categories: background excluding $\Xi^{0}$ or $\bar{\Xi}^{0}$ particles (\textit{i.e.}, non-$\Xi^{0}\bar{\Xi}^{0}$ background) and background involving $\Xi^{0} \bar{\Xi}^{0}$. The non-$\Xi^{0}\bar{\Xi}^{0}$ background is estimated via the two-dimensional sideband analysis of the data samples, as shown in Fig.~\ref{fig:sideband_B_C}. The signal region is defined as 1.296~GeV/$c^2 < M_{\pi^0\Lambda/\pi^0\bar{\Lambda}} <$ 1.331~GeV/$c^2$ and the lower and upper sideband regions are defined as 1.251~GeV/$c^2 < M_{\pi^0\Lambda/\pi^0\bar{\Lambda}} <$ 1.286~GeV/$c^2$ and 1.341~GeV/$c^2 < M_{\pi^0\Lambda/\pi^0\bar{\Lambda}} <$ 1.376~GeV/$c^2$, respectively, shown in Fig.~\ref{fig:sideband_B_C}. The sideband regions are positioned approximately $\pm 2 \sigma$ away from the signal region, where $\sigma$ corresponds to the mass resolution of $\Xi^0$. We divide the sideband regions into two categories: region A and region B as shown in Fig.~\ref{fig:sideband_B_C}. The number of background events $N_{\rm bkg}$ is determined by $0.5N_{A}-0.25N_B$, where $N_A$ and $N_B$ are the sum of all events in the regions A and B, respectively. From this method, the number of non-$\Xi^{0} \bar{\Xi}^{0}$ background events is estimated to be $1032\pm 30$ and fixed in the following fit. 

The background involving $\Xi^{0}\bar{\Xi}^{0}$ comprises $J/\psi \to \Xi^{0} \bar{\Xi}^{*0}$, $J/\psi \to \Xi^{*0} \bar{\Xi}^{0}$, and $J/\psi \to \Xi^{0} \bar{\Xi}^{0}$, which have fake or missing photon.  The shapes and detection efficiencies of these background channels are obtained from MC simulation, and their event counts are estimated using the branching fractions from the PDG~\cite{PDG}. Especially, to reduce the large systematic uncertainties from the poorly measured branching fraction of the $J/\psi \to \Xi^{0} \bar{\Xi}^{*0}$ ($J/\psi \to \Xi^{*0} \bar{\Xi}^{0}$) channel~\cite{DM2:1987nsg}, we estimate its contribution using the precisely measured isospin partner channel, $J/\psi \to \Xi^{*-}\bar{\Xi}^+$~\cite{BESIII:2019jot}. The number of $J/\psi \to \Xi^0\bar{\Xi}^{*0}$ and $J/\psi \to \Xi^{*0}\bar{\Xi}^0$ events is estimated to be $294 \pm 7$ each. Additionally, the number of $J/\psi \to \Xi^0\bar{\Xi}^0$ background events is estimated to be $158 \pm 5$. The uncertainties here are attributed to the uncertainties of the branching fractions.

\section{\boldmath Signal extractions and branching fraction measurement}
\label{sec:bf}

Signal yield is extracted from an unbinned maximum-likelihood fit to the distribution of $\Xi^0\bar{\Xi}^0$ invariant mass. Accounting for the interference between $J/\psi \to \gamma\eta_c\to\gamma\Xi^{0} \bar{\Xi}^{0}$ and $J/\psi \to\gamma \Xi^{0} \bar{\Xi}^{0}|_{\rm{non}-resonance}$, where $J/\psi \to\gamma \Xi^{0} \bar{\Xi}^{0}|_{\rm{non}-resonance}$ corresponds to the direct $J/\psi \to\gamma \Xi^{0} \bar{\Xi}^{0}$ decay bypassing the $\eta_c$ resonance state, the fitting probability density function (PDF) as a function of mass (m) reads:
\begin{equation*}
F(m) =  \left[ \varepsilon(m) \left | e^{i\phi} {\cal S}(m) + \alpha \sqrt{{\cal N}(m)} \right|^{2} \right] \otimes \sigma_{\rm res} + {\cal B}(m),
\label{eq:fit}
\end{equation*}
where ${\cal S}(m)$, ${\cal N}(m)$, and ${\cal B}(m)$ represent the probability density functions (PDFs) of the signal, the $J/\psi \to\gamma \Xi^{0} \bar{\Xi}^{0}|_{\rm{non}-resonance}$, and the combined background, respectively; ${\cal N}(m)$ is modeled as a 2nd-order polynomial with its shape fixed from MC study and the detail of combined background ${\cal B}(m)$ can be seen in Sec.~\ref{sec:background_study}; $\varepsilon(m)$ is the mass-dependent efficiency modeled as a piecewise function according to the PHSP MC sample; $\sigma_{\rm res}$ represents the detector resolution function modeled as a double Gaussian function with the parameters 
derived from the signal MC sample; The interference phase $\phi$ and the strength of
the non-resonance component $\alpha$ are allowed to vary in the fit. In addition, ${\cal S}(m)$ reads
\begin{equation*}
{\cal S}(m) =  \sqrt{E_{\gamma}^{3} \times {\rm damp}(E_{\gamma})} \times {\cal BW}(m),
\label{eq:fit2}
\end{equation*}
where 
$E_{\gamma}$ is the radiated photon energy
with $E_{\gamma} = \frac{m^{2}_{J/\psi}-m^{2}}{2m_{J/\psi}}$, $m_{J/\psi}$ here represents $J/\psi$ nominal mass.
$E_{\gamma}^{3}$ reflects the energy dependence of the magnetic-dipole  transition matrix element; damp($E_{\gamma}$) describes a function to damp the diverging tail caused by the $E_{\gamma}^{3}$ dependence and is given in the form of $\frac{E_0^2}{E_{\gamma}E_0+(E_{\gamma}-E_0)^2}$ as used by KEDR~\cite{Anashin:2010nr}. The equation $E_0=\frac{m^2_{J/\psi}-m^2_{\eta_c}}{2m_{J/\psi}}$ denotes the peaking energy of the transition photon, where $m_{\eta_c}$ represents $\eta_c$ nominal mass; The ${\cal BW}(m)= \frac{\Gamma}{(m^2_{\eta_c} - m^2 - i m_{\eta_c} \Gamma)}$~\cite{Castro:2024prg} is the signal component described by the Breit-Wigner function, where $\Gamma$ is the $\eta_c$ decay width and is set to the value in PDG~\cite{PDG}.

Figure~\ref{fig:conclusion_result} shows the $M(\Xi^0\bar{\Xi}^0)$ distribution and the best fitting results. The solutions are listed in Table~\ref{tab:sum_fit}, with two fitting results corresponding to the destructive interference and constructive interference, where $N_{\rm obs}$ is true signal yield and $N_{\rm sig}$ is the total yields of the coherent signal component containing the true signal, the non-resonance contribution, and their interference term. 
To estimate the statistical uncertainties, we generate 5000 sets of fitting parameters through multivariate Gaussian sampling with the covariance matrix from the nominal fit, and then corresponding 5000 sets of true signal event counts are calculated. These true signal event counts follow a Gaussian distribution, and its relative standard deviation value represents the statistical uncertainties, which are 2.64\% and 2.63\% for the destructive and constructive solutions, respectively.  The $\chi^2$/ndf~\cite{PDG} value of the fit is $108.68/76=1.4$, where ndf is the number of degrees of freedom. The statistical significance of the $\eta_c$ signal is larger than 30$\sigma$. 

The product of branching fractions of $J/\psi\to\gamma\eta_c$ and $\eta_c\to\Xi^{0} \bar{\Xi}^{0}$ is calculated with
\begin{equation*}
\mathcal{B}(J/\psi\to\gamma\eta_c) \times \mathcal{B}(\eta_c \to \Xi^0 \bar{\Xi}^0) = \frac{N_{\rm obs}}{N_{J/\psi} \times \mathcal{B}_{\rm sub} \times \varepsilon},
\end{equation*}
where $N_{J/\psi}$ is the total number of $J/\psi$  events in data~\cite{BESIII:2021cxx}, $N_{\rm obs}$ is the number of true signal events,
$\mathcal{B}_{\rm sub}=\mathcal{B}^2(\Xi^0 \to \pi^0 \Lambda)\times \mathcal{B}^2(\Lambda\to p \pi^-)\times \mathcal{B}^2(\pi^0\to\gamma\gamma)$, and $\varepsilon=1.78\%$ is the detection efficiency. The results are summarized in Table~\ref{tab:sum_fit}.

\begin{table}[hpbt]
\centering
\caption{Summary of the fit results for the $M(\Xi^0\bar{\Xi}^0)$ distribution. The two columns correspond to the solutions with destructive and constructive interference between the signal and the non-resonance process $J/\psi \to\gamma \Xi^{0}\bar{\Xi}^{0}|_{\rm{non}-resonance}$, respectively, where the uncertainties are statistical.} 
\label{tab:sum_fit}
\begin{tabular}{ccc}
\hline \hline 
Results & Destructive & Contructive \\\hline
$N_{\rm sig }$ & $2202\pm62~~~$ & $2205\pm62~~~$\\
$N_{\rm obs }$ & $1629\pm43~~~$ & $1330\pm35~~~$\\
$\alpha$ & $0.024\pm0.002$ & $0.028\pm0.002$\\
$\phi$(rad) & $0.39\pm0.37$ & $6.13\pm0.12$ \\
$\mathcal{B}(J/\psi\to\gamma\eta_c\to\gamma\Xi^0\bar{\Xi}^0)(10^{-5})$ & $2.30\pm0.06$ & $1.88\pm0.05$ \\
 \hline \hline
\end{tabular}
\end{table}

\section{\boldmath Systematic uncertainties}

\begin{table}[tbp]
  \caption{Systematic uncertainties (\%) in the branching fraction measurements of $\mathcal{B}(J/\psi\to\gamma\eta_c\to\gamma\Xi^0\bar{\Xi}^0)$, where a dash (-) indicates that the systematic effect is negligible.}\label{tab:tot_sys}
  \centering
  \small
  \begin{tabular}{ccc}
    \toprule
        \hline
       \hline
    Source & Destructive & Constructive \\
    \midrule
     \hline
    Photon reconstruction & 0.5 & 0.5 \\
    $\Xi^0/\bar\Xi^0$ reconstruction & 5.6 & 5.6 \\
    Kinematic fit & - & - \\
    Damp function & 2.3 & 3.0 \\
    $\eta_{c}$ width & 0.9 & 1.0 \\
    Polarization of $\Lambda$ and $\Xi^0$ & 3.4 & 3.4 \\
    Detector resolution & 0.5 & 0.4 \\
    Efficiency curve & 0.3 & 0.2 \\
    Shape of $J/\psi \to \gamma \Xi^{0} \bar{\Xi}^{0}$ & 3.0 & 3.6 \\
    Number of $J/\psi \to \Xi^{0} \bar{\Xi}^{0}$ & 0.7 & 0.6 \\
    Number of $J/\psi \to \Xi^{*0} \bar{\Xi}^{0}(\Xi^{0} \bar{\Xi}^{*0})$ & 1.2 & 1.1 \\
    Non-$\Xi^0\bar\Xi^0$ Background & 1.7 & 3.0 \\
    $\mathcal{B}(\Lambda\to p\pi^-)$ & 1.6 & 1.6 \\
    Number of $J/\psi$ events & 0.5 & 0.5 \\
    \midrule
      \hline
    Total & 8.2 & 9.0 \\
    \bottomrule
          \hline
         \hline
  \end{tabular}
\end{table}

Table~\ref{tab:tot_sys} summarizes the sources of systematic uncertainties, which are described in details as follows.

\begin{itemize}
\item[(i)] \textit{Photon reconstruction.}  Using the control sample of $J/\psi\to\gamma\mu^+\mu^-$, the systematic uncertainty of photon reconstruction is estimated to be 0.5\%.

\item[(ii)] \textit{$\Xi^0/\bar\Xi^0$ reconstruction.} The $\Xi^0$ reconstruction efficiency, which includes
the $\pi^0$ reconstruction, $\Lambda$ reconstruction, and $\Xi^0/\bar \Xi^0$ mass window, is studied with the control sample $\psi(3686)\to\Xi^0\bar\Xi^0$. The efficiency difference between data and MC samples is taken as the systematic uncertainty, with a value of 5.6\%.

\item[(iii)] \textit{Kinematic fit.}  The helix parameter correction in the 4C kinematic fit is applied in the nominal result. To estimate the uncertainty associated with the helix  parameter correction, half of the difference in efficiency with and without the helix parameter correction is considered as the systematic uncertainties, following the method described in Ref.~\cite{BESIII:2012mpj}.

\item[(iv)] \textit{Damp function.} An alternative damping function used by the CLEO Collaboration~\cite{CLEO:2008pln},  damp$(E_{\gamma})=\exp(-E_{\gamma}^2/8\beta^2)$ where $\beta$ is fixed to be 0.065 GeV, is chosen to estimate the related uncertainty.
The difference using the two damping functions is taken as the systematic uncertainty. 

\item[(v)] \textit{$\eta_{c}$ width.}  The systematic uncertainty caused by the $\eta_{c}$ width in the fit is evaluated by varying the width within one standard deviation~\cite{PDG}, and taking the result with the highest relative systematic uncertainty. 

\item[(vi)] \textit{Polarizations of $\Lambda$ and $\Xi^0$.} We consider the extreme case that the $\Lambda$ and $\Xi^0$ polarizations are equal to $+1$ and $-1$, and take the largest relative difference of efficiency compared to the nominal values as the systematic uncertainty.

\item[(vii)] \textit{Detector resolution.} The systematic uncertainty due to resolution is estimated by convolving an additional small Gaussian function in the fit. The relative difference between the fit with and without this convolution is taken as the systematic uncertainty.

\item[(viii)] \textit{Efficiency curve.} The systematic uncertainty of the efficiency curve is evaluated by comparing two different functional forms. The nominal form is modeled as a piecewise function consisting of a third-order polynomial followed by a first-order polynomial. For an alternative form, it is changed to a 4th-order polynomial followed by a 2nd-order polynomial. The relative difference between the two fits is considered as the systematic uncertainty. 

\item[(ix)] \textit{Shape of $J/\psi \to \gamma \Xi^{0} \bar{\Xi}^{0}$.} To estimate the systematic uncertainty from the shape of $J/\psi \to \gamma \Xi^{0} \bar{\Xi}^{0}$, we generate 5000 sets of parameters describing the $J/\psi \to \gamma \Xi^{0} \bar{\Xi}^{0}$ shape through multivariate Gaussian sampling with the covariance matrix from the fitting of ${\cal N}(m)$. Using these parameters as input, we recalculate the true signal yield, and the relative standard deviation of the recalculated true signal yield values is taken as the systematic uncertainty.

\item[(x)] \textit{Number of $J/\psi \to \Xi^{0} \bar{\Xi}^{0}$ background events.} The systematic uncertainty from the fixed  $J/\psi \to \Xi^{0} \bar{\Xi}^{0}$ background yield is estimated by modifying the number by $\pm 1\sigma$, where $\sigma$ is derived from the branching fractions and the statistical fluctuations of the $J/\psi \to \Xi^{0} \bar{\Xi}^{0}$ event counts. The relative difference is taken as the systematic uncertainty.

\item[(xi)] \textit{Number of $J/\psi \to \Xi^{*0} \bar{\Xi}^{0}(\Xi^{0} \bar{\Xi}^{*0})$ background events.}
Due to isospin conservation, we utilize $J/\psi \to \Xi^{*-}\bar{\Xi}^+$ to estimate the contribution of $J/\psi \to \Xi^{*0} \bar{\Xi}^{0}$ ($J/\psi \to \Xi^{0} \bar{\Xi}^{*0}$). The branching ratio of $J/\psi \to \Xi^{*-}\bar{\Xi}^+$ is $(3.18\pm0.08)\times 10^{-4}$~\cite{BESIII:2019jot}, while that of $J/\psi \to \Xi^{*0} \bar{\Xi}^{0}$ ($J/\psi \to \Xi^{0} \bar{\Xi}^{*0}$) is $(3.2\pm1.4)\times 10^{-4}$~\cite{DM2:1987nsg}. The uncertainty of $J/\psi \to \Xi^{*0} \bar{\Xi}^{0}$ ($J/\psi \to \Xi^{0} \bar{\Xi}^{*0}$) is significantly larger compared to $J/\psi \to \Xi^{*-}\bar{\Xi}^+$. This effectively reduces the systematic uncertainty. The systematic uncertainty from the fixed $J/\psi \to \Xi^{*0} \bar{\Xi}^{0}$ ($J/\psi \to \Xi^{0} \bar{\Xi}^{*0}$) background yield is estimated by modifying the number by $\pm 1\sigma$ of $J/\psi \to \Xi^{*-}\bar{\Xi}^+$, similarly to $J/\psi \to \Xi^{0} \bar{\Xi}^{0}$. 

\item[(xii)] \textit{Non-$\Xi^0\bar\Xi^0$ Background.} The systematic uncertainty from the non-$\Xi^0\bar\Xi^0$ background is estimated by changing the sideband regions.  The difference of the fitted signal yield is taken as the systematic uncertainty. 

\item[(xiii)] \textit{Branching fraction.} The uncertainty of the
branching fraction of $\mathcal{B}(\Lambda\to p\pi^-)$ contributes 0.8\%~\cite{PDG}.

\item[(xiv)] \textit{Total number of $J/\psi$ events.} The uncertainty of the total number of $J/\psi$ events is 0.44\%~\cite{BESIII:2021cxx}. 

\end{itemize}

\section{\boldmath Summary}

Using $(10087\pm 44)\times 10^6$ $J/\psi$  events collected with the BESIII detector, we report the first observation of the decay $\eta_{c} \to \Xi^{0} \bar{\Xi}^{0}$. The analysis accounts for interference effects with the irreducible non-resonant background.
The product branching fraction for the full decay chain $J/\psi\to\gamma\eta_c\to\gamma\Xi^0\bar{\Xi}^0$ is measured to be 

\begin{equation*}
\begin{aligned}
    \mathcal{B}(J&/\psi\to\gamma\eta_c\to\gamma\Xi^0\bar{\Xi}^0)_{\rm{destructive}} \\ =& (2.30 \pm 0.06_{\rm stat.} \pm 0.19_{\rm syst.}) \times 10^{-5},
\end{aligned}
\end{equation*}

\begin{equation*}
\begin{aligned}
    \mathcal{B}(J&/\psi\to\gamma\eta_c\to\gamma\Xi^0\bar{\Xi}^0)_{\rm{constructive}} \\ =& (1.88 \pm 0.05_{\rm stat.} \pm 0.17_{\rm syst.}) \times 10^{-5}.
\end{aligned}
\end{equation*}
Normalizing with the $\mathcal{B}(J/\psi \rightarrow \gamma \eta_{c})=(1.41\pm0.14)\times 10^{-3}$~\cite{PDG}, the branching fraction of the decay $\eta_c \to \Xi^0 \bar{\Xi}^0$ is determined to be

\begin{equation*}
\begin{aligned}
    \mathcal{B}(\eta&_c\to\Xi^0\bar{\Xi}^0)_{\rm{destructive}} \\ =& (1.63 \pm 0.04_{\rm stat.} \pm 0.21_{\rm syst.}) \times 10^{-3},
\end{aligned}
\end{equation*}

\begin{equation*}
\begin{aligned}
    \mathcal{B}(\eta&_c\to\Xi^0\bar{\Xi}^0)_{\rm{constructive}} \\ =& (1.33 \pm 0.03_{\rm stat.} \pm 0.18_{\rm syst.}) \times 10^{-3}.
\end{aligned}
\end{equation*}
These results are significantly larger than the expectations from the pQCD helicity selection rule, confirming that this process is not forbidden and providing an important new constraint on charmonium baryonic decays. The branching-fraction ratios $\mathcal{B}(\eta_c \to \Xi^0\bar{\Xi}^0)/\mathcal{B}(\eta_c \to \Xi^-\bar{\Xi}^+)$ are determined to be $(1.56\pm0.29)$ and $(1.51\pm0.32)$ for the destructive and constructive solutions, respectively, and are consistent within $2\sigma$ with isospin symmetry expectations~\cite{BESIII:2012gcb}. This discrepancy may arise from the fact that previous measurements did not take into account the interference between the $\eta_c$ state and non-resonant contributions. Meanwhile, the measured branching fraction is compatible with the IML model prediction of $(0.63\sim1.25) \times 10^{-3}$~\cite{Liu:2010um} and is substantially larger than the quark--diquark model prediction of $\mathcal{B}(\eta_c \to \Xi^0 \bar{\Xi}^0) = 1.46 \times 10^{-4}$~\cite{Anselmino:1987du}. These results provide a valuable experimental benchmark for theoretical studies of helicity selection rule violation and charmonium decays into baryon--antibaryon final states.

\section{\boldmath ACKNOWLEDGMENTS}
The BESIII Collaboration thanks the staff of BEPCII (https://cstr.cn/31109.02.BEPC) and the IHEP computing center for their strong support. This work is supported in part by National Key R\&D Program of China under Contracts Nos. 2025YFA1613900, 2023YFA1606000, 2023YFA1606704; National Natural Science Foundation of China (NSFC) under Contracts Nos. 12375070, 11635010, 11935015, 11935016, 11935018, 12025502, 12035009, 12035013, 12061131003, 12192260, 12192261, 12192262, 12192263, 12192264, 12192265, 12221005, 12225509, 12235017, 12342502, 12361141819; the Chinese Academy of Sciences (CAS) Large-Scale Scientific Facility Program; the Strategic Priority Research Program of Chinese Academy of Sciences under Contract No. XDA0480600; CAS under Contract No. YSBR-101; 100 Talents Program of CAS; Shanghai Leading Talent Program of Eastern Talent Plan under Contract No. JLH5913002; Shanghai Top Talent Program of Eastern Talent Plan under Contract No. BJZH2025073; The Institute of Nuclear and Particle Physics (INPAC) and Shanghai Key Laboratory for Particle Physics and Cosmology; ERC under Contract No. 758462; German Research Foundation DFG under Contract No. FOR5327; Istituto Nazionale di Fisica Nucleare, Italy; Knut and Alice Wallenberg Foundation under Contracts Nos. 2021.0174, 2021.0299, 2023.0315; Ministry of Development of Turkey under Contract No. DPT2006K-120470; National Research Foundation of Korea under Contract No. NRF-2022R1A2C1092335; National Science and Technology fund of Mongolia; Polish National Science Centre under Contract No. 2024/53/B/ST2/00975; STFC (United Kingdom); Swedish Research Council under Contract No. 2019.04595; U. S. Department of Energy under Contract No. DE-FG02-05ER41374

\end{document}